\documentclass[twocolumn,showpacs,prl]{revtex4}

\usepackage{graphicx}%
\usepackage{dcolumn}
\usepackage{amsmath}
\usepackage{latexsym}

\begin {document}

\title {Quasistatic Scale-free Networks}

\author {G. Mukherjee$^{1,2}$ and S. S. Manna$^1$}
\affiliation 
{
$^1$Satyendra Nath Bose National Centre for Basic Sciences 
    Block-JD, Sector-III, Salt Lake, Kolkata 700098, India \\
$^2$Bidhan Chandra College, Asansol 713304, 
                  Dt. Burdwan, West Bengal, India
}

\begin{abstract}
A network is formed using the $N$ sites of an one-dimensional lattice
in the shape of a ring as nodes and each node with the initial degree
$k_{in}=2$. $N$ links are then introduced to this 
network, each link starts from a distinct node, the other end being connected to any
other node with degree $k$ randomly selected with an attachment probability 
proportional to $k^{\alpha}$. Tuning the control parameter $\alpha$ we
observe a transition where the average degree of the largest node $\langle k_m(\alpha,N) \rangle$
changes its variation from $N^0$ to $N$ at a specific transition point of $\alpha_c$. 
The network is scale-free i.e., the nodal degree distribution has a
power law decay for $\alpha \ge \alpha_c$.
\end{abstract}
\maketitle

      The nodal degree distribution function of a scale-free network (SFN) has a power 
   law tail \cite {barabasi}. Empirical data obtained for several social, biological 
   and computational networks have confirmed existence of such probability distributions 
   decaying as power laws \cite {albert,neural,collab}. For example the World-wide web \cite {web} 
   which is a network of web pages and the hyper-links among various pages and the Internet 
   network \cite {Faloutsos} of routers or autonomous systems follow power laws as: 
   $P(k) \sim k^{-\gamma}$. The exponent $\gamma$ varies between 2 and 3 for these 
   networks.

      Networks are classified here as `growing', `quasistatic' and `static'.
   In a growing network (GN) the nodes are introduced in the network one 
   after another. After introducing the node a link is introduced connecting this 
   node to one of its previous nodes. Therefore, in a GN both the numbers of nodes
   as well as links grow with time. In contrast, in the case of a 
   quasistatic network (QN) a fixed number $N$ of nodes are present at the initial stage. 
   $N$ Links are then introduced one after another between pairs of nodes
   using some specific probability distribution. Therefore in a QN, 
   the number of nodes in the network is fixed but the number of links grow with time.
   Finally, in a static network (SN) both the number of nodes as well as the
   number of links remain fixed and donot grow with time.

      Barab\'asi and Albert (BA) proposed \cite {barabasi} a simple model for a growing SFN 
   that has the following two essential ingredients, namely: (i) A network grows 
   from an initial set of $m_o$ nodes with $m < m_o$ links among them. Further, 
   at every time step a new node is introduced and is randomly connected to $m$ 
   previous nodes. (ii) Any of these $m$ links of the new node introduced at time 
   $t$ connects a previous node $i$ with an attachment probability $\pi_i(t)$ 
   which is linearly proportional to the degree $k_i(t)$ of the $i$-th node at 
   time $t$: $\pi^{BA}_i(t) \sim k_i(t)$. For BA model $\gamma=3$ \cite {albert}. 

%---------------------------------------------------------------------------
\begin{figure}[t]
\begin{center}
\includegraphics[width=7cm]{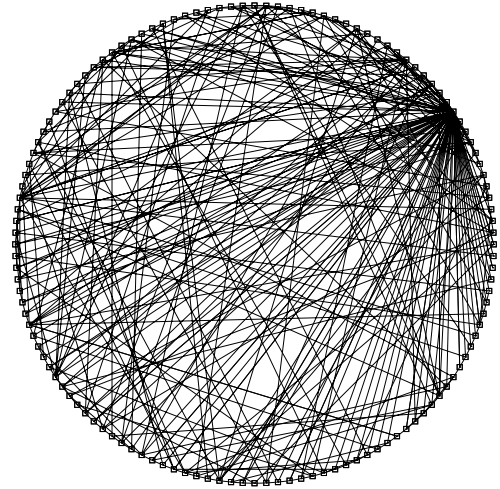}
\end{center}
\caption{
A quasistatic scale-free network with 128 nodes placed on a circular ring 
so that the initial degree $k_{init}$ of each node is equal to 2. Nodes are then selected 
one after another without repetition and are randomly linked to other nodes with 
an attachment probability proportional to $k^{\alpha}$. In this picture $\alpha=2.6$
is used and a large degree node is visible.
}
\end{figure}
%---------------------------------------------------------------------------

      The second criterion reflects the
   phenomenon of `rich gets richer' i.e., a node with a large degree attracts more
   nodes to get linked. Krapivsky et. al. showed that this linear dependence is a 
   necessary condition and any other non-linear dependence on the degree
   destroys the scale-free behaviour i.e., the power law variation of the degree
   distribution \cite {redner}. Indeed, if in general the degree dependence has a variation with the 
   $\alpha$-th power of the degree, it has been shown that the scale-free nature of 
   the BA model network exists only for $\alpha =1$ and for no other value of it, 
   smaller or larger.

      Several other interesting networks are also studied. Erd\"os and R\'enyi
   studied long ago random graphs of $N$ nodes where links are introduced with 
   probability $p$ between arbitrary pairs of randomly selected nodes \cite 
   {random}. A largest component of such a network connecting nodes of the order
   of $N$ appears at a particular value of $p_c=1/N$. Very recently SFNs
   are studied on Euclidean spaces where the BA attachment probability is 
   modified by a link length $\ell$ dependent factor \cite 
   {Manna_Sen,euclid,jost,yook2}. Load distribution in Scale-free networks
   has also been studied \cite {Goh}.

      In this paper we ask the question if the growing condition of the
   BA model is really a necessity to achieve a scale-free network. We will see
   below that it is not, a suitable choice of the attachment probability
   in a quasistatic network may result a power law decay for the degree distribution as well. We call
   our model as the Quasitatic Scale-free Network (QSFN). Recently Doye has
   shown that the network topology of a potential energy landscape is a
   static scale-free network \cite {Doye}. Assigning a quenched intrinsic fitness
   to every node and using the attachment probabilities depending on the
   fitnesses it is also possible to get SFNs without growth and preferential
   attachments \cite {Caldarelli}. A steady state model for scale-free graphs 
   has also been proposed \cite {Eppstein}.

%---------------------------------------------------------------------------
\begin{figure}[t]
\begin{center}
\includegraphics[width=7cm]{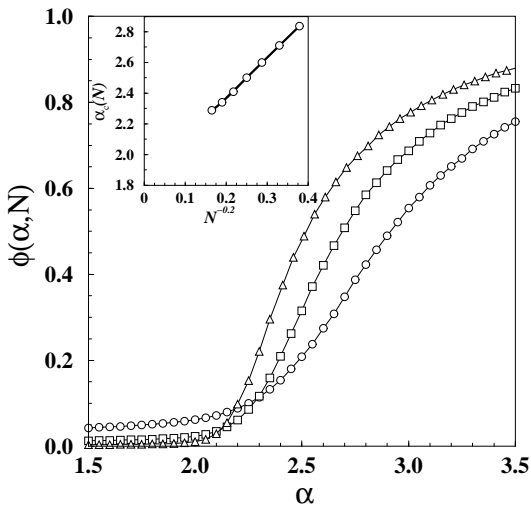}
\end{center}
\caption{
The variation of the order parameter $\phi(\alpha,N)$ which is the average degree
of the largest node divided by $N$ with $\alpha$ for three different network sizes
$N=2^8$ (circle), $2^{10}$ (square) and $2^{12}$ (triangle). The inset shows the
plot of the threshold values $\alpha_c(N)$ with $N^{-0.2}$ for seven different $N$
values from $2^7$ to $2^{13}$ increasing by a factor of 2 at each step.
}
\end{figure}
%---------------------------------------------------------------------------

      Our quasistatic model starts with an one-dimensional regular lattice in the form 
   of a circular ring as in the Watts and Strogatz's Small-world Network \cite {WS}. 
   The lattice has $N$ nodes serially marked from $i$=1 to $N$ and $N$ links between
   successive pairs of nodes with periodic boundary condition. Each node is therefore
   connected to only two of its nearest neighbours situated on the opposite sides and 
   the degree $k_i$ for each node $i$ is exactly 2 to begin with. We add to this 
   system another $N$ distinct links in total, such that a new link starts from each 
   node. A $t$-th link is added at time $t$ and one end of it is attached to the $t$-th
   node, the other end of the link is connected to a node $j$ with degree $k_j(t)$ 
   selected randomly from the rest of the $N-1$ nodes of the system using the following 
   attachment probability:
\begin {equation}
\pi_j(t) \sim k_j^\alpha(t)
\end {equation}
   where $\alpha$ is a continuously tunable parameter. Therefore when the network is 
   complete no node is left with degree 2.

      At this point we would like to clearly distinguish between our QSFN and the
   model B \cite {modelB}. Model B also starts with a collection of $N$ nodes but
   no links, so that the initial degree of each node is strictly zero and the 
   graph has $N$ components each with only one node. At each time step a node $i$ 
   is selected randomly with uniform probability and is connected to another node 
   $j$ with a probability $\pi_j(t) \sim k_j(t)$. This implies that isolated nodes
   are being linked one by one to a single connected component of the graph. For 
   this component, both the number of nodes as well as the links grow with time 
   and therefore model B is clearly a growing network. This is in contrast to our 
   QSFN model where the initial network is a $N$ node connected graph which grows 
   by adding further links but not the nodes any more.

%---------------------------------------------------------------------------
\begin{figure}[t]
\begin{center}
\includegraphics[width=7cm]{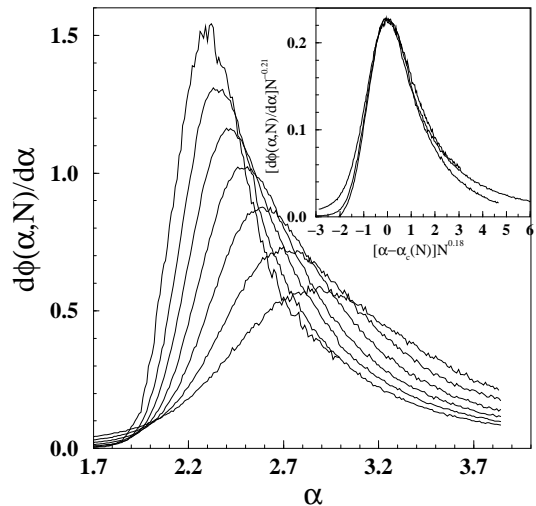}
\end{center}
\caption{
The local derivative of the order parameter $d\phi(\alpha,N)/d\alpha$
has been plotted with $\alpha$ for 7 different
$N$ values starting from $2^7$ and increasing by a factor of 2. Scaling of
the derivative is done in the inset
for $N=2^8, 2^{10}$ and $2^{12}$ by plotting
$[d\phi(\alpha,N)/d\alpha] N^{-0.21}$ vs. $[\alpha-\alpha_c(N)]N^{0.18}$.
}
\end{figure}
%---------------------------------------------------------------------------

      In the case of $\alpha=0$ in QSFN, all nodes have equal probabilities to get 
   connected by a new link and this is the random network model \cite {random} for 
   which the degree distribution is the Poisson distribution \cite {albert}. For 
   $\alpha > 0$, a node with larger degree has a higher probability to get connected 
   by links. The case of $\alpha=1$ corresponds to the similar attachment
   probability as in the BA model of SFN. Our simulation results show that the degree distribution in 
   this case has an exponentially decaying form: $P(k) \sim \exp(-ak)$ with 
   $a \approx 1.08$. On continuously increasing $\alpha$ further 
   larger degree nodes become more probable and we see that the degree distribution 
   decays less sharply and
   changes to a stretched exponential form as: $P(k) \sim \exp[-k^{\chi(\alpha)})]$. 
   The exponent $\chi(\alpha)$ decreases continuously from its value 1 at $\alpha=1$
   to $\chi \approx 0.3$ at $\alpha=1.75$.

      On increasing the parameter $\alpha$ further the system makes a transition
   to a different behaviour where a single node having the maximum degree $k_m$ 
   connects to a finite fraction of the $N$ nodes. However this transition takes 
   place at a specific value $\alpha_c$ of $\alpha$ so that the degree distribution 
   has a power law tail and therefore the network is scale-free for all $\alpha \ge \alpha_c$.

      Naturally the control parameter in this model is $\alpha$ where as like the 
   phenomenon of percolation, an order parameter in this problem may be the average 
   degree of the largest node $\phi(\alpha,N) = \langle k_m(\alpha,N) \rangle/N$. 
   We monitor the variation of $\phi(\alpha,N)$ with $\alpha$ and plot them in Fig. 2 
   for networks of three different sizes $N=2^8,2^{10}$ and $2^{12}$. A sharp
   increase in the order parameter is observed around $\alpha=2$, again the sharpness 
   of the curves increases with $N$. The first derivative $d\phi(\alpha,N)/d\alpha$ 
   of the order parameter is plotted in Fig. 3 for seven different network sizes. 
   Every curve has a peak at some $N$ dependent value of $\alpha_c(N)$ where 
   $\phi(\alpha,N)$ increases at the fastest rate. For locating the precise value 
   of $\alpha_c(\infty)$ where this transition is taking place for an infinitely 
   large network, we plot $\alpha_c(N)$ values as a function of $N^{-1/\nu}$ in the 
   inset of Fig. 2. Using a trial value of $\nu=5$ we could get all seven points on 
   a straight line and on extrapolating to the $N \rightarrow \infty$ limit we get 
   $\alpha_c(\infty) = 1.85 \pm 0.10$.

%---------------------------------------------------------------------------
\begin{figure}[t]
\begin{center}
\includegraphics[width=7cm]{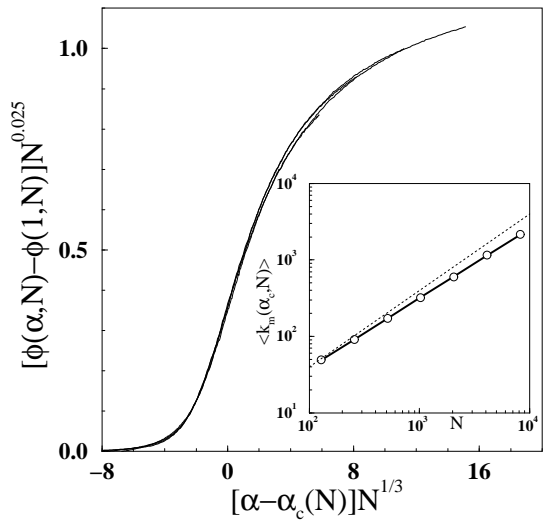}
\end{center}
\caption{
Scaling of the order parameter:
$[\phi(\alpha,N)-\phi(1,N)]N^{0.025}$ is plotted with $[\alpha - \alpha_c(N)]N^{1/3}$
for the data collapse. The inset shows that average degree of the largest node
$\langle k_m(\alpha,N) \rangle$ varies as $N^{\mu}$ at the transition point
$\alpha_c$ where we obtained $\mu = 0.91 \pm 0.04$. The dotted line, having a slope
unity, is given to compare with the data.
}
\end{figure}
%---------------------------------------------------------------------------

      All the first derivative curves are suitably scaled in the inset of Fig. 3
   and the data are collapsed. Each curve is first shifted by an amount
   $\alpha_c(N)$. Then the abscissa is scaled by $N^{0.18}$ and the ordinate
   is scaled by $N^{-0.21}$ and we get a nice collapse of three curves
   of sizes $N=2^8, 2^{10}$ and $2^{12}$. This analysis implies that the
   order parameter curves in Fig. 2 become progressively sharper as the
   network size $N$ gradually increases and $\phi(\alpha,N)$ increases very
   rapidly at $\alpha_c(\infty)$. This variation is very similar to the
   variation of the order parameter (the fraction of mass in the largest
   cluster) in the percolation phenomena. 

      The scaling of the order parameter $\phi(\alpha,N)$ is shown in Fig. 4. 
   Asymptotically as $N \rightarrow \infty$ the $\phi(\alpha,N) \rightarrow 0$ 
   for $\alpha < \alpha_c$ but for finite $N$ we subtract the $\phi(1,N)$
   from $\phi(\alpha,N)$ and try to scale only the difference. We assume 
   the following scaling behaviour:
\begin {equation}
[\phi(\alpha,N)-\phi(1,N)]N^{\beta/\nu} = {\cal F} [(\alpha-\alpha_c)N^{1/\nu}]
\end {equation}
   where the scaling function ${\cal F}(x) \rightarrow x^{\beta}$ for $x << 1$.
   From Fig. 4 we get $\nu=3$ and $\beta=0.075$. This value of $\nu$ is not very
   consistent with its value 5 obtained in the inset of Fig. 2. We believe 
   this difference is due to finite size of our simulations.

      The dependence of the average degree $\langle k_m(\alpha_c,N) \rangle $
   of the largest node right at the transition point $\alpha = \alpha_c$      
   is also studied and plotted in the inset of Fig. 4 on a double logarithmic scale.
   Assuming a power law dependence on the network size with an exponent $\mu$ as
   $\langle k_m(\alpha_c,N) \rangle \sim N^{\mu}$ the plot gives a value of 
   $\mu = 0.91 \pm 0.04$. Here $\mu$ is an exponent similar to the
   fractal dimension of the infinite incipient percolation clusters at the
   percolation threshold.
   
%---------------------------------------------------------------------------
\begin{figure}[t]
\begin{center}
\includegraphics[width=7cm]{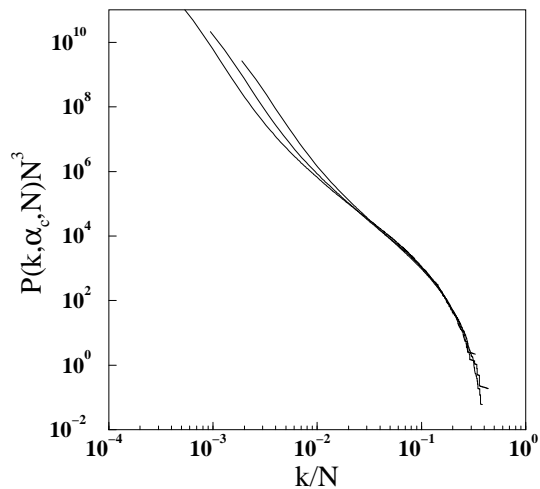}
\end{center}
\caption{
Scaled degree distributions at the transition points $\alpha_c(N)$ for different
network sizes have been plotted. Data collapse is obtained for large $k$ values when $P(k,\alpha_c,N)N^{\eta}$
has been plotted with $k/N^{\xi}$ with $\eta=3$ and $\xi=1$. This gives the
degree distribution exponent $\gamma=\eta/\xi=3$.
}
\end{figure}
%---------------------------------------------------------------------------

Finally the degree distribution $P(k,\alpha_c(N),N)$ is calculated for different
network sizes $N$ at their transition points $\alpha_c(N)$.
In these calculations we have not counted the maximum degree node
similar to what is usually done to estimate the cluster size distribution
at the percolation threshold.
Except for very small $k$ values all distribution curves have straight portions
on double logarithmic plots and the region over which nearly constant slopes
are observed, increases with increasing $N$. The fluctuating data
is suitably binned: for each $k$ value the data is averaged over the
bin from $k$ to $2k-1$ and is plotted at $[k(2k-1)]^{1/2}$.
We assume a scaling form like:
\begin {equation}
P(k,\alpha_c(N),N)N^{\eta} \sim {\cal G}(k/N^{\xi})
\end {equation}
where ${\cal G}(x)$ is expected to be an universal scaling function
such that ${\cal G}(x) \rightarrow x^{-\gamma}$ for $x << 1$ 
so that $\gamma = \eta/\xi$. The scaling plot is shown in Fig. 5
where we get $\eta=3$ and $\xi=1$ giving $\gamma=3$ as in
the BA model of SFN. This is a very interesting result that even on
a static network with a suitable attachment probability $k^{\alpha}$
with a non-trivial value of $\alpha$ one gets a scale-free degree
distribution for the network, the distribution being the same as the
BA model of SFN.

      What happens when $\alpha > \alpha_c$? Our observation is that
   degree distribution is still a power law but the exponent $\gamma$
   is now $\alpha$ dependent. In the range of $\alpha \ge 3$, the
   degree exponent $\gamma$ is approximately equal to $\alpha$ whereas
   for $\alpha_c\le\alpha\le3$, $\gamma(\alpha) \approx 3$.

      We also have tried a more stochastic version of this model where
   to add a link, we select in parallel two nodes arbitrarily  from the 
   whole set of $N$ nodes using the attachment probability in Eqn. (1). 
   If these two nodes are not the same node, we connect them. This is 
   also a QSFN but the difference here is both the nodes of each link 
   are selected randomly where as in the QSFN described above only one 
   node is selected randomly and the other node is selected systematically 
   one after another from the set of nodes. Our numerical studies indicate 
   that behaviours of both versions are very similar.

      Another variation of our model has been studied with initial degrees
   of all nodes as $k_{in}=1$ which corresponds to a situation where alternate
   pairs of nodes on an one-dimensional ring shaped lattice are linked.
   Again we numerically observe that the $\alpha_c = 1.67 \pm 0.10$ and
   $\gamma(\alpha_c) \approx 2.5$ for this model.
   
   To summarize, we have studied a quasistatic scale-free network (QSFN)
where we have a collection of $N$ nodes initially present, each node
is having $k_{in}$ initial degree. A system of $N$ links are then
introduced, one corresponding to each node. That means we systematically
attach the one end of each link to one node, the other end of the
link is probabilistically attached to any other node of degree $k$ with a
probability proportional to $k^{\alpha}$. Our numerical study indicates that there 
exists a transition point $\alpha_c$ beyond which the resulting network 
has a scale-free structure so that the degree distribution has a power law
tail for $\alpha \ge \alpha_c$.

      We thankfully acknowledge P. Sen, D. Stauffer and D. Dhar for critical reading of the
   manuscript and many useful comments and also A.-L. Barab\'asi for
   helpful suggestions. GM gratefully acknowledges the hospitality at the
   S. N. Bose National Centre for basic sciences.

\leftline {Correspondence to: manna@boson.bose.res.in}


\begin{thebibliography}{90}

\bibitem {barabasi} A.-L. Barab\'asi and R. Albert, Science, {\bf 286}, 509 (1999).

\bibitem {albert} R. Albert and A.-L. Barab\'asi, Rev. Mod. Phys. {\bf 74}, 47 (2002).

\bibitem {neural} J. J. Hopfield and A. V. M. Herz,
Proc. Natl. Acad. Sci. USA, {\bf 92}, 6655 (1995).

\bibitem {collab} M. E. J. Newman, Proc. Nat. Acad. Sci. USA, {\bf 98}, 404 (2001);
                  arXiv:cond-mat/0011155.

\bibitem {web} S. Lawrence and C. L. Giles, Science, {\bf 280}, 98 (1998);
               Nature, {\bf 400}, 107 (1999), R. Albert, H. Jeong and A.-L. Barab\'asi,
               Nature, {\bf 401}, 130 (1999).

\bibitem {Faloutsos} M. Faloutsos, P. Faloutsos and C. Faloutsos, Proc.
               ACM SIGCOMM, Comput. Commun. Rev., {\bf 29}, 251 (1999).

\bibitem {redner} P. L. Krapivsky, G. J. Rodgers and S. Redner, Phys. Rev. Lett.
{\bf 86}, 5401 (2001); P. L. Krapivsky and S. Redner, Phys. Rev. E, {\bf 63}, 066123 (2001).

\bibitem {random} P. Erd\"os and A. R\'enyi, Publ. Math. Debrecen, {\bf 6}, 290 (1959).

\bibitem {Manna_Sen} S. S. Manna and P. Sen, arXiv:cond-mat/0203216.

\bibitem{euclid}  S. Jespersen and A. Blumen, Phys. Rev. E {\bf 62}, 6270 (2000);
J. Kleinberg, Nature {\bf 406}, 845 (2000); S. N. Dorogovtsev, J.F.F. Mendes and A.N.Samukhin,
arXiv:cond-mat/0206467.

\bibitem {jost} J. Jost and M. P. Joy, arXiv:cond-mat/0202343.

\bibitem {yook2} S. H. Yook, H. Jeong, A.-L. Barab\'asi and Y. Tu,
Phys. Rev. Lett. {\bf 86}, 5835 (2001).

\bibitem {Goh} K.-I. Goh, B. Khang and D. Kim, Phys. Rev. Lett. {\bf 87}, 278701 (2001).

\bibitem {Doye} J. P. K. Doye, Phys. Rev. Lett. {\bf 23}, 238701 (2002).

\bibitem {Caldarelli} G. Caldarelli, A. Capocci, P. De Los Rios and M. A. Mu\~noz, arXiv:cond-mat/0207366.

\bibitem {Eppstein} D. Eppstein and J. Wang, arXiv:cs.DM/0204001.

\bibitem {WS} D. J. Watts and S. H. Strogatz, Nature, {\bf 393}, 440 {1998};
D. J. Watts, {\it Small Worlds: The Dynamics of Networks Between order and Randomness},
(Princeton 1999).

\bibitem {modelB} A.-L. Barab\'asi, R. Albert and H. Jeong, Physica A, {\bf 272}, 173 (1999).
\end{thebibliography}
\end {document}